\documentclass[12pt]{iopart}
\usepackage[dvips]{graphicx}

\begin{document}
\title{Electronic detection of charged particle effects in a Penning trap}
\author{D F A Winters, M Vogel, D M Segal and R C Thompson}
\address{Blackett Laboratory, Imperial College London, Prince Consort Road, London SW7 2BW, United Kingdom}
\ead{d.winters@imperial.ac.uk}
\begin{abstract}
We present a thorough analysis of the electronic detection of charged particles, confined in a Penning trap, via image charges induced in the trap electrodes. Trapping of charged particles in an electrode structure leads to frequency shifts, which are due to image charge and space charge effects. These effects are of importance for Penning trap experiments which involve high charge densities or require high precision in the motional frequencies. Our analysis of image charges shows that only (higher order) odd powers of the particle displacement lead to induced charge differences, giving rise to a signal. This implies that, besides the centre-of-mass frequency of a trapped particle cloud, also higher order individual particle frequencies induce a signal, which can be picked up by an electronic detection circuit attached to the trap electrodes. We also derive analytic expressions for the image charge and space charge induced frequency shifts and perform simulations of space charge effects. In relation to this, we discuss the consequences of the shifted particle frequencies for resistive cooling of the particle motion.
\end{abstract}
\submitto{\jpb}
\maketitle

\section{Introduction}
Penning traps have proven to be versatile tools for extended investigations of localised particles under well defined conditions, thus allowing for high-precision experiments. Such experiments often rely on precise particle manipulation and require knowledge of the motional frequencies and other dynamic properties of the trapped particles. However, the presence of charged particles in a Penning trap is accompanied by image charge and space charge effects, which change the effective trapping potential and therefore lead to oscillation frequency changes. Image charge effects are caused by particle-induced image charges in the trap electrodes and in most cases lead to only small frequency shifts. However, in precision experiments they still need to be considered \cite{BRA99,FAR95}. Space charge effects are caused by the charge density of trapped particles and lead to considerable frequency shifts and broadening \cite{BRA99,FAR95,VAN89,DHO95,THO04,VAN98}.

In the following we discuss ideal particle motions in a Penning trap. We present a thorough analysis of the formation of image charges and determine image charge and space charge induced frequency shifts of the particle motions. Electronic detection of the axial particle frequency (section 5) will be discussed, followed by a section on the implications for resistive cooling of the (shifted) axial frequency (section 6). The aim of this study is to give a consistent quantitative treatment of these effects. Similar effects also occur in other types of traps, such as Paul or radio-frequency traps, but their analysis is beyond the scope of this study.

\section{Single particle motion in a Penning trap}
In a Penning trap, axial ($z$) confinement of a single particle is assured by an electrode structure, typically one ring electrode and two endcaps, that creates a static electric quadrupole field at the centre of the trap. Radial ($xy$) confinement is obtained by a static magnetic field $B$ along the trap axis ($z$). Once trapped, a charged particle oscillates between the two endcaps with `axial frequency' $\omega_z$. In the radial plane there are two motions with distinct frequencies: a fast motion with `modified cyclotron frequency' $\omega_+$, and a slow motion with `magnetron frequency' $\omega_-$. The latter represents a slow drift about the centre of the trap. The sum of these two frequencies equals the `cyclotron frequency' $\omega_c=qB/m$, where $q$ is the charge of the single particle and $m$ its mass.

The electric field inside the Penning trap is described by the quadrupole potential $U$, which is given by
\begin{equation} 
U=\frac{U_0}{R_0^2} \left(2z^2 - r^2 \right)
\end{equation}
where $U_0$ is the trapping potential. The constant $R_0^2=r_0^2 + 2z_0^2$ takes the geometry of the electrodes into account, where $z_0$ and $r_0$ are the distances from the trap centre to the endcap and the ring, respectively. A convenient choice for the trap dimensions is $r_0^2=2 z_0^2$, so that the trap depth is $U_0/2$ \cite{BRO86}.

In the axial or $z$-direction, the equation of motion is
\begin{equation} 
m \ddot{z}=-q \frac{\partial U}{\partial z}=-\frac{4qU_0}{R_0^2}z
\end{equation}
From equation (2) the axial frequency can be directly obtained, {\it i.e.}
\begin{equation} 
\omega_z^2=\frac{4qU_0}{mR_0^2}
\end{equation}

In the radial or $xy$-plane, the equations of motion are coupled by the magnetic field, {\it i.e.}
\begin{eqnarray} 
m\ddot{x} &=& q \left[\frac{2U_0}{R_0^2}x + \dot{y}B \right] \nonumber\\
m\ddot{y} &=& q \left[\frac{2U_0}{R_0^2}y - \dot{x}B \right]
\end{eqnarray}

It can be shown that the modified cyclotron frequency $\omega_+$ and the magnetron frequency $\omega_-$ are given by \cite{GHO95}
\begin{equation} 
\omega_{\pm}= \frac{\omega_c}{2} \pm \sqrt{ \left( \frac{\omega_c}{2} \right)^2 - \frac{1}{2} \omega^2_z }
\end{equation}
A more detailed discussion of single particle motion in an ideal Penning trap is given in \cite{BRO86,GHO95,KRE91}.

\section{Image charge effects}
The above description only holds true for the idealised case where there are no interactions between the particle and its environment. Therefore we need to consider the interaction between a single charged particle and the `image charges' induced in the trap electrodes (this section and \cite{TIN01}). When there are more particles in the trap, also particle-particle interactions or `space charge effects' have to be considered (section 4). In order to understand the formation of image charges we will study two different geometries: a charge between two parallel conducting plates, and a charge between two hollow conducting spheres. It is worth pointing out that electronic detection and resistive cooling are based on the detection of image charges via a resonant RLC-circuit attached to the trap electrodes (sections 5 and 6).

\subsection{Charge between two parallel conducting plates}
The interaction between a charged particle and a single conducting plate can be described via the `method of images' \cite{FEY65,JAC75}. A particle with real charge $q$, positioned at a distance $z$ in front of the surface of an infinite conducting plate, induces a surface charge density $\sigma(r,z)$ which can be calculated by Gauss' law and is given by
\begin{equation} 
\sigma(r,z)=-\frac{q}{2 \pi} \frac{z}{\left( r^2+z^2 \right)^{3/2} }
\end{equation}
where $r^2=x^2+y^2$ and $(x,y)$ is the position on the surface. On an infinite plate the charge induced within a radius $R$ is given by \cite{BLE65}
\begin{equation} 
q_i=\int_0^R \sigma(r,z) 2\pi r dr = -q \left( 1-\frac{z}{\sqrt{R^2+z^2}} \right)
\end{equation}
such that for $R \rightarrow \infty$ the induced charge is $q_i=-q$.

\begin{figure}[!t]
\begin{center}
\centering
\includegraphics[width=7.5cm]{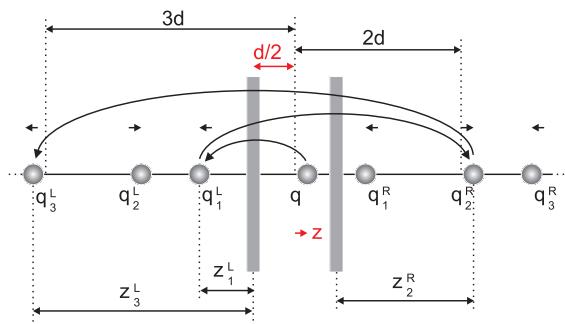}
\caption{A real charge between two parallel conducting plates and its induced images. The arrows illustrate the successively induced images $q_i$ at distances $z_i$ from the plates, $R$ is the plate radius and $d$ the plate separation.}
\label{fig1}
\end{center}
\end{figure}

A charged particle located between two parallel conducting plates induces a series of images. This is schematically shown in figure~\ref{fig1} by the curved arrows. Upon a displacement $z$, the real charge $q$ induces an image charge $q_1^L$ in the left plate at $z_1^L$. This, in turn, induces an image $q_2^R$ in the right plate at $z_2^R$, and so on. A second sequence of images also appears, starting with $q_1^R$ induced in the right plate. For a positively charged particle ($q>0$), positive (even) images are located at even multiples of the plate separation $d$, {\it i.e.} the distances $\pm 2d,\pm 4d,\ldots$. Negative (odd) images are located at odd multiples of $d$, {\it i.e.} the distances $\pm d,\pm 3d,\ldots$. The top dashed lines indicate the positions of the image charges for $z=0$, the lower ones (and the small black arrows) for $z>0$.

From equation (7) it is clear that the image charge depends on the radius $R$ of the plates, which implies that for finite $R$ the image $q_i$ is `imperfect'. For an image $i=1,2,\ldots$ induced in the left plate, the position $z_i^L$ and the corresponding image charge $q_i^L$ are given by
\begin{eqnarray} 
z_i^L &=& (2i-1)\frac{d}{2}-z(-1)^i \quad \textrm{and} \nonumber\\
q_i^L &=&-q_{i-1}^R \left( 1-\frac{z_i^L}{\sqrt{R^2+(z_i^L)^2}} \right)
\end{eqnarray}
where $q_0^R=q_0^L=q$.
The expressions for the images induced in the right plate are similar, but with the sign of $z$ reversed. Here we have made the assumption that the charge induced in a finite plate of radius $R$ is the same as that induced within a radius $R$ of an infinite plate. When $R$ is small compared to $d$, the images are imperfect and the series terminates rapidly. The induced charge difference $\Delta q$ between the two plates is defined as $\Delta q = q^L - q^R$, where $q^L$ and $q^R$ are the sums of the charges induced in the left and in the right plate, respectively. When the particle is located close to the right plate, $\Delta q$ is positive, and vice versa. For a finite plate radius $R<d$, it suffices to consider only 3 images induced in both plates, as shown in figure~\ref{fig1}, to obtain a good description of $\Delta q$. For such a simple case, it can be shown that $\Delta q$ is given by a power series with only odd powers of $z$, {\it i.e.}
\begin{equation} 
\Delta q \approx \gamma_1 z + \gamma_3 z^3 + \gamma_5 z^5 + \ldots
\end{equation}
The expressions for $\gamma_i$ show that mainly the odd images contribute to $\Delta q$ and that they lead to (higher order) odd powers of $z$. For increasing $R$, the variation becomes close to linear, which can be verified by extending the series to infinity. In that case the higher order odd terms vanish almost completely, leaving $\Delta q \approx \gamma_1 z$.

\begin{figure}[!t]
\begin{center}
\centering
\includegraphics[width=7.5cm]{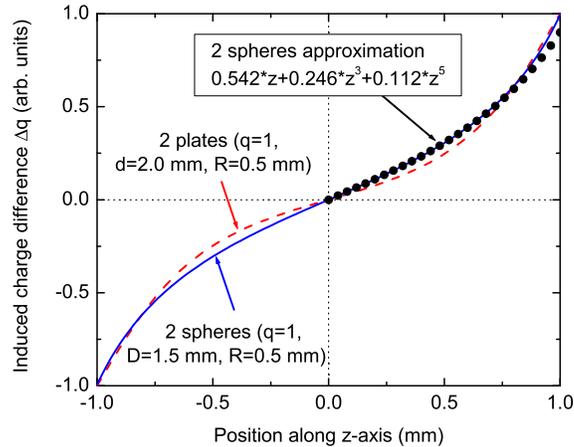}
\caption{Induced charge difference $\Delta q$ plotted versus the position of the real charge $q$ centered between: a) two parallel conducting plates (dashed line), and b) two hollow conducting spheres (solid line). The black dots are calculated using equation (9) for $z>0$.}
\label{fig2}
\end{center}
\end{figure}

In figure~\ref{fig2} the calculated dependence of $\Delta q$ is plotted as a function of the position $z$ of the real charge $q=1$ between two parallel conducting plates (dashed line) with radius $R=0.5$ mm, which are separated by a distance $d=2.0$ mm. Figure~\ref{fig2} clearly shows that for small $z$, $\Delta q$ is almost linear with $z$, whereas for larger $z$, $\Delta q$ varies with odd powers of $z$. For fixed $R$, when $d \approx R$, $\Delta q$ is very non-linear at large $z$, due to strong higher order terms. When $d > R$, $\Delta q$ is rather linear, but the induced charge difference is small. For infinitely large plates $R \rightarrow \infty$, the response is very linear. Below we will show that higher order odd terms not only occur for planar electrodes, but also for curved electrodes (solid line in figure~\ref{fig2}).

\subsection{Charge between two hollow conducting spheres}
\begin{figure}[!t]
\begin{center}
\centering
\includegraphics[width=7.5cm]{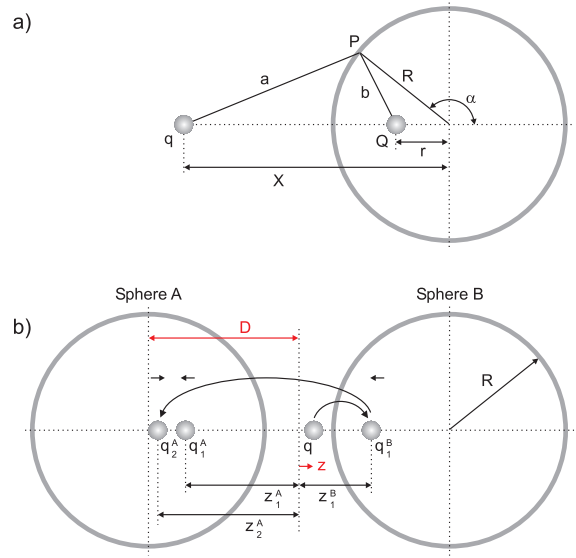}
\caption{a) Schematic of a hollow conducting sphere used to obtain the relationship between the real charge $q$ and its image $Q$. b) A real charge $q$ between two hollow conducting spheres and the corresponding first image charges $q_1^A$, $q_1^B$ and $q_2^A$.}
\label{fig3}
\end{center}
\end{figure}

In a conventional Penning trap geometry, the endcap electrodes are better approximated by two hollow conducting spheres, ignoring the presence of the ring electrode. The case of a charge outside a single hollow conducting sphere can easily be solved (see {\it e.g.} \cite{BLE65}). As illustrated in figure~\ref{fig3}a), the real charge $q$, located outside the sphere at $X$, induces an image charge $Q$ located at $r$ inside the sphere.

The potential at a point $P$ on the surface of the sphere, is given by \cite{BLE65}
\begin{equation} 
V=\frac{1}{4 \pi \epsilon_0} \left[ \frac{Q}{a}+\frac{q}{b} \right]
\end{equation}
where the distances
\begin{eqnarray} 
a^2 &=& R^2+X^2+2XR\cos(\alpha) \nonumber\\
b^2 &=& R^2+r^2+2rR \cos(\alpha)
\end{eqnarray}
are given by the cosine rule, and $\epsilon_0$ is the permittivity of free space. After satisfying the boundary condition that the potential at any point $P$ on the grounded conducting surface is zero, the real charge and its image are related as follows
\begin{eqnarray} 
r &=& R^2/X \nonumber\\
q &=& -QR/X
\end{eqnarray}

Now two spheres $A$ and $B$ with radius $R$ and separated by $2D$, as depicted in figure \ref{fig3}b), are considered. Upon a displacement $z$, the real charge $q$ induces an image charge $q_1^B$ in sphere $B$ at $z_1^B$. This, in turn, induces an image $q_2^A$ at $z_2^A$ in sphere $A$, and so on. A second sequence of images also appears, starting with $q_1^A$ induced in sphere $A$. For the image $i$ (integer) induced in sphere $A$, the position $z_i^A$ and the corresponding image charge $q_i^A$ are given by
\begin{eqnarray} 
z_i^A &=& D-\frac{R^2}{(D-z_{i-1}^A)} \quad \textrm{and} \nonumber\\
q_i^A &=& \frac{-q_{i-1}^B R}{(D-z_i^B)}
\end{eqnarray}
where $z_0^B=z_0^A=z$ and $q_0^B=q_0^A=q$. The expressions for the images induced in sphere $B$ are similar, but with the sign of $z$ reversed. In analogy to the case of parallel conducting plates, the induced charge difference is defined as $\Delta q = q^A - q^B$, where $q^A$ and $q^B$ are the sums of the charges induced in sphere $A$ and sphere $B$, respectively. Also in this case, for a finite sphere radius $R$, it suffices to consider only 3 images in both spheres to obtain a good approximation to $\Delta q$. Again, for such a simple case, it can be shown that $\Delta q$ is given by a power series with only odd powers of $z$.

In figure~\ref{fig2} $\Delta q$ is plotted as a function of the position $z$ of the real charge $q=1$ between two hollow conducting spheres (solid line) with radius $R=0.5$ mm, which are separated by a distance $2D=3.0$ mm, giving a gap of $d=2.0$ mm. The black dots are calculated using equation (9) for the 2 spheres case, taking only 3 images in both spheres into account. From equation (9) in figure~\ref{fig2} it can clearly be seen that the higher order odd terms are non-negligible for larger values of $z$. This analysis shows that also here, as in the case of 2 plates, mainly the odd images contribute to $\Delta q$ and that they again lead to higher order odd powers of $z$. When $D \approx R$, $\Delta q$ is very non-linear at large $z$, due to strong higher order terms, but when $D > R$, $\Delta q$ is rather linear.

\section{Frequency shifts due to image charge and space charge effects}
The image charge and space charge effects change the effective trapping potential and therefore lead to frequency changes. In the following, starting from the potentials at the trap centre, we derive analytic expressions for these frequency shifts and compare their predictions with experimental results obtained in measurements of clouds of ions confined in a Penning trap. In the case of the space charge effects we have also performed simulations, which confirm and support our analysis and are in line with experimental results.

\subsection{Image charge shift}
As can be seen in figure~\ref{fig1}, the effective distance between the particle and the positive (even) images does not change as it moves between the plates. Therefore, only negative (odd) images exert a net force on the particle. For a single confined charged particle, the total force is thus given by the sum of the force due to the quadrupole potential and the image force. From figure~\ref{fig1}, it can be seen that the image force is given by
\begin{equation} 
F^{im}_z=\frac{-q^2}{4 \pi \epsilon_0} \sum_{i} \left[ \left( \frac{1}{id-2z} \right)^2 - \left( \frac{1}{id+2z} \right)^2 \right]
\end{equation}
where the summation is over odd images $i$ only. Wineland {\it et al.} \cite{WIN75} stated that the image force is 
given by $F^{im}_z \approx -8q^2z/(4 \pi \epsilon_0 d^3)$, under the assumption that the displacement $z$ is much smaller than the separation between the plates $d$. Our analysis shows that equation (14) is better approximated, even for small $z$, by including the higher order terms, {\it i.e.}
\begin{equation} 
F^{im}_z=\frac{-q^2}{4 \pi \epsilon_0} \left[ \frac{8z}{d^3} \left(\frac{1}{1^3}+ \frac{1}{3^3}+\frac{1}{5^3} +\dots \right)\right]
\end{equation}
Obviously, for larger amplitudes, {\it i.e.} $z \approx d/2$, this approximation is no longer accurate. In general, equation (14) is much better approximated, and over a larger range of $z$-values, when represented by a power series expansion. Ignoring the small higher order terms in equation (15), the image force is given by
\begin{equation} 
F^{im}_z=\frac{-q^2}{4 \pi \epsilon_0} \left[ \frac{8z}{d^3}+ \frac{64 z^3}{d^5}+\dots \right]
\end{equation}
In analogy with section 3, it can be shown that mainly odd images contribute to the image force and that they lead to (higher order) odd powers of the particle displacement $z$. However, since the higher order effects are small for harmonic motion (small $z$), we will here just consider the first order term in equation (16). The corresponding equation of motion in the axial direction is given by
\begin{equation} 
m \ddot{z}=-\frac{4qU_0}{R_0^2}z+\frac{2q^2}{\pi \epsilon_0 d^3}z
\end{equation}
The image charged shifted axial frequency is thus
\begin{equation} 
\label{eq19}
\omega''^2_z=\frac{4qU_0}{mR_0^2}-\frac{2q^2}{\pi \epsilon_0m d^3}
\end{equation}
To first order, the shift itself is given by
\begin{equation} 
\Delta \omega_z = \omega''_z - \omega_z \approx -\frac{1}{\pi\epsilon_0}\frac{q^2}{m \omega_z d^3}
\end{equation}
where the higher order effects have been neglected. If there are more particles between the plates, the image force and thus the axial frequency shift scales with the number of particles $N$. A result similar to equation (19) is obtained when the particle is considered to move inside a hollow conducting sphere \cite{VAN89,POR01}. The image force affects the centre-of-mass(c.m.) motion of a particle cloud, because it is an induced external force acting on the cloud as a whole.

Due to its radial motions, {\it i.e.} the magnetron and modified cyclotron motions, the charged particle also induces image charges in the ring electrode, which affect the radial frequencies. In order to calculate this radial image force, an approximate treatment similar to that given in section 3.2 can be applied, where the ring electrode is a small central section of the surface of a sphere. 

As can be seen from figure~\ref{fig3}a), the force between the charges $q$ and $Q$ is given by
\begin{equation} 
F^{im}_r=\frac{1}{4 \pi \epsilon_0} \frac{q Q}{(X-r)^2}
\end{equation}
By substituting equations (12) into (20) one obtains
\begin{equation} 
F^{im}_r=\frac{-q^2}{4 \pi \epsilon_0} \frac{R}{(R^2-r^2)^2} r \qquad (r \leq R)
\end{equation}
which is the radial image force on the particle due to one sphere. However, now the real charge $q$ is located inside the sphere and its image lies outside. As the ring is considered as a small central section of the sphere's surface, for this case there will also be an image force like equation (21), albeit lower in magnitude due to the fact that not all the field lines originating from the charge $q$ are directed at the section. Or, in other words, the image charge that is formed is imperfect. A charged particle with, for example, a magnetron radius $r$ will thus experience a shifted magnetron frequency due to the induced charge in the ring. 

For a positively charged particle, the radial image force is directed radially outward, following the direction of the electric quadrupole force. The corresponding shifted radial frequencies are given by
\begin{equation} 
\omega''_{\pm}= \frac{\omega_c}{2} \pm \sqrt{ \left( \frac{\omega_c}{2} \right)^2 - \frac{1}{2} \omega^2_z - \gamma \frac{F^{im}_r}{m}}
\end{equation}
where $\gamma$ is related to the ratio of the surface areas of the ring and the sphere.
 
The image charge shift is only significant for very small traps and in most cases represents much less than 1 \% of the axial frequency. Measurements of the image charge shift dependence on the number of particles have been performed for $^1$H$^+$, $^2$H$^+$, $^3$He$^+$, $^3$He$^{2+}$, and $^{12}$C$^{4+}$ ions \cite{VAN89} and frequency shifts were observed even for small numbers of particles. The magnetron frequency was shifted upwards and the modified cyclotron frequency was shifted downwards, as expected from equation (22). The relative magnitudes of the observed shifts, about 20 mHz per ion per charge, are also in fair agreement with those predicted by equation (22). For the cyclotron frequency $\omega_c=\omega_++\omega_-$ no shift has been observed, as expected.

\subsection{Space charge shift}
Similarly to the image charge effect, there is a potential due to the presence of space charge which adds to the trapping potential and thus changes the trapping frequencies. The space charge potential $U'$ is the potential seen by a particle due to the presence of all other particles and can be derived in a manner similar to the quadrupole potential. As a simple example, it is assumed that the particle cloud can be described by a perfect sphere with 
radius $R'$ and homogeneous charge density $\rho=N q/V'$, where $N$ is the number of particles and $V'=4 \pi R'^3/3$ the volume of the sphere. Laplace's law in spherical coordinates ($\eta,\theta,\phi$) applied to this particle cloud states that
\begin{equation} 
\frac{\partial^2 U'}{\partial \eta^2}+\frac{2}{\eta}\frac{\partial U'}{\partial \eta} =-\frac{\rho}{\epsilon_0}
\end{equation}
where spherical symmetry with respect to $\theta$ and $\phi$ holds. 

Outside the sphere the potential can be described by that of a point charge, {\it i.e.}
\begin{equation} 
U'(\eta \geq R')=\frac{N q}{4 \pi \epsilon_0 \eta}=\frac{\rho R'^3}{3 \epsilon_0 \eta}
\end{equation}
Inside the sphere the potential should vary with $\eta^2$ in order to satisfy Laplace's equation. The final form of the potential is
\begin{equation} 
U'=\frac{\rho}{6 \epsilon_0} (3R'^2-\eta^2)
\end{equation}

For a single particle, the equation of motion in the $z$-direction, with the substitution $\eta \rightarrow z$, becomes
\begin{equation} 
m \ddot{z}=-q \frac{\partial}{\partial z} \left( U+U' \right)=-\frac{4qU_0}{R_0^2}z + \frac{\rho q}{3 \epsilon_0}z
\end{equation}
Both terms on the right-hand side of equation (26) vary linearly with $z$, {\it i.e.} equation (26) has the same form as equations (2) and (17). Therefore the space charge shifted axial frequency is given by (see {\it e.g.} \cite{JEF83})
\begin{equation} 
\omega'^2_z =\frac{4qU_0}{m R_0^2} - \frac{\rho q}{3 m \epsilon_0} = \omega_z^2-\frac{\omega_p^2}{3}
\end{equation}
where $\omega_p^2=\rho q/(m \epsilon_0)$ is the so-called `plasma frequency' (see {\it e.g.} \cite{MAJ04}). The factor 3, which appears in the denominator of the space charge term in equation (27), accounts for the geometry of the particle cloud. In a more general case, when the particle cloud is not a sphere but rather an ellipsoid, this constant changes. If the cloud shape is `prolate' the factor is smaller than 3, if the shape is `oblate' it is larger \cite{JEF83}. To first order, the absolute frequency shift is given by 
\begin{equation} 
\Delta \omega_z = \omega_z'- \omega_z \approx - \frac{n q^2}{6 m \epsilon_0 \omega_z}
\end{equation}
where we have used the particle number density $n=N/V'=\rho/q$. The relative frequency shift
\begin{equation} 
\frac{\Delta \omega_z}{\omega_z} \equiv \frac{\omega_z'- \omega_z}{\omega_z} \approx - \frac{n q R_0^2}{24 \epsilon_0 U_0}
\end{equation}
depends linearly on the charge density $\rho=n q$ and thus, for a given cloud size, on the total number of ions $N=\rho V'$.

The space charge limit is reached for $|U'| \geq |U|$, {\it i.e.} when the space charge potential compensates the trapping potential. In this case there is no potential minimum and confinement is lost.

If the space charge density is not homogeneous, but has a Maxwell-Boltzmann distribution, the space charge shift depends on the average distance of the particle from the cloud centre \cite{MEI88}. This leads to a distribution of shifted axial frequencies. However, the c.m.-motion of a particle cloud is not affected by intra-cloud interactions and thus remains unshifted by space charge effects \cite{WIN75}. A similar broadening of the axial frequency distribution occurs when the trapping potential is anharmonic. In that case, the axial frequency depends on the amplitude of the motion such that a distribution of particle energies automatically leads to a distribution of axial frequencies. To correct for such anharmonicities, most Penning traps have compensation electrodes (see {\it e.g.} \cite{GAB89}).

The space charge potential $U'$ also affects the radial motions and thus the radial frequencies of the particle. Due to the difference in signs between the $z$- and $r$- dependences in equation (1), the space charge term now has the opposite sign. This can be shown via the substitution $U \rightarrow U' +U$ in equation (2). The space charge shifted radial frequencies are given by
\begin{equation} 
\omega'_{\pm}= \frac{\omega_c}{2} \pm \sqrt{ \left( \frac{\omega_c}{2} \right)^2 - \frac{1}{2} \omega^2_z - \frac{\omega^2_p}{3} }
\end{equation}
Thus, there is a downwards shift of the modified cyclotron frequency $\omega_+$ and an upwards shift of the magnetron frequency $\omega_-$ with increasing charge density $\rho$. 

Measurements of the magnetron frequency of an ion cloud in a Penning trap with $r_0=5$ mm, $z_0=3.5$ mm, $U_0\approx 6$ V and $B=1$ T were performed by Dholakia {\it et al.} \cite{DHO95} through laser cooling and laser spectroscopic techniques. For a single ion they yield a magnetron frequency of $\sim 40$ kHz, while for clouds with ion number densities of the order of $10^7$ cm$^{-3}$, the corresponding value is shifted to $\sim 60$ kHz. This is in fair accordance with the behaviour expected from equation (30).

\subsection{Simulations of the space charge shift}
In an effort to check the validity of the results of the calculation, the influence of space charge has been studied using SIMION, which allows for charged particle tracking in static electric and magnetic fields created in a user-defined geometry. The simulations have been performed for hydrogen-like lead ions \cite{WIN05,VOG05}($m=207$ u and $q=81$) confined in a cylindrical open endcap Penning trap \cite{GAB89}.

The ion cloud is simulated by a cubic array, typically of size 5x5x5, in which a charge of magnitude $e'$ is assigned to each array point. So, for a cloud of $N=125$ charged particles $e'=q \times e$ ($e=1.6 \times 10^{-19}$ C), whereas for a cloud of $N=125000$ particles, $e'=q \times 1000 \times e$. The particle-particle interactions are simulated by SIMION using the grouped Coulomb repulsion $e'$. Within this approximation, small clouds are clearly better described than large ones. It may be speculated that a simulation where $N$ particles are simulated by $N$ array points would result in the frequency shift expected from equation (29). This, however, is not within our computational capabilities since the computation time scales with $N^2$. Once a simulation produced several periods of oscillation (for all particles), the simulation was terminated and the particle frequencies were extracted from the data by means of Fourier analysis.

\begin{figure}[!t]
\begin{center}
\centering
\includegraphics[width=7.5cm]{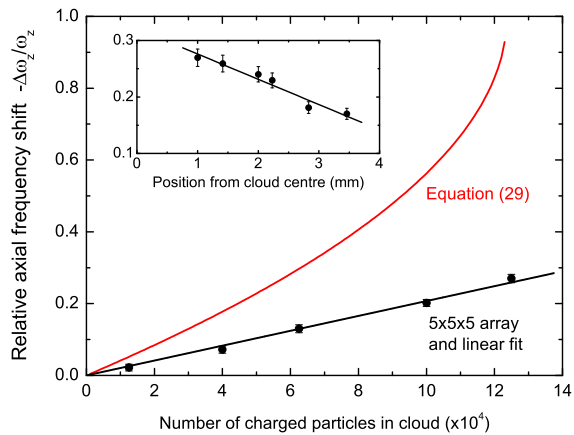}
\caption{Simulated relative axial frequency shift $-\Delta \omega_z / \omega_z$ plotted versus the number of particles $N$. Equation (29) is also shown for comparison. The inset shows the dependence of the shift on the distance from the cloud centre.}
\label{fig4}
\end{center}
\end{figure}

The results of our simulations are shown in figure~\ref{fig4}, where the relative axial frequency shift is plotted versus the number of particles in the cloud. The parameters were: $z_0=15$ mm, $U_0=1000$ V, $B=1$ T, $N=125000$ (5x5x5 array) and $n=4 \times 10^6$ m$^{-3}$. The simulations show a linear relationship between the relative axial frequency shift and the number of particles. Also shown is a plot of equation (29), which behaves linearly for small $n$ ($\propto N$) as expected, but clearly deviates from the results of our simulations for larger numbers of particles. This discrepancy is attributed to the limited number of array points used in our simulations, the cubic lattice geometry (rather than spherical), and the method used to simulate high ion densities ({\it i.e.} via $e'$). However, for large $N$ also the approximation used for the derivation of equation (29) is no longer accurate.

The c.m.-motion of the cloud is found to be invariant under the influence of interaction between the particles, as expected. However, as shown by the inset in figure~\ref{fig4}, the relative frequency shift is larger for particles that are initially positioned close the cloud centre than for particles that are further out. This is attributed to the fact that the amount of space charge surrounding the particle drops with increasing distance from the cloud centre.

\section{Electronic detection of the axial motion}
In order to experimentally determine the axial frequency, a resonant RLC-circuit with a high quality factor (see section 6) is connected to the endcaps. A Fourier analysis of the induced current $I$ in the RLC-circuit, picked up by a sensitive amplifier, then yields $\omega_z$. For the case of a charge between two parallel conducting plates, the current flow at any time is
\begin{equation} 
I=\frac{\partial \Delta q}{\partial t} = \frac{\partial \Delta q}{\partial z} \frac{\partial z}{\partial t}=v_z \frac{\partial \Delta q}{\partial z}
\end{equation}
where $\Delta q$ is the induced charge difference between the two endcaps. For parallel plates with large radii, $\Delta q$ varies linearly with $z$, {\it i.e.} $\Delta q = \gamma_1 z$. However, in many other cases, as we have shown above in equation (9) and figure~\ref{fig2}, there can be higher order odd terms as well. For example, if $\Delta q = \gamma_1 z + \gamma_3 z^3$, it can be shown that the electronic circuit also picks up the higher order term. For simple harmonic motion, of the form $z=z_0 \cos(\omega_z t)$, one obtains
\begin{eqnarray} 
\Delta q &=& \gamma_1 z_0 \cos (\omega_z t)\\ \nonumber
&+& \frac{\gamma_3 z_0^3}{4} \left[ 3 \cos \left( \omega_z t \right) + \cos \left( 3 \omega_z t \right) \right] \nonumber
\end{eqnarray} 
The cubic term therefore induces signals at $\omega_z$ and $3 \omega_z$.

Consider a single particle with mass $m$ and charge $q$ which is moving inside a particle cloud of mass $M$ and charge $Q$. In this case only the c.m.-motion of the cloud will normally be detected. If the particles are all identical, then $q/m=Q/M$. If the position of $q$ is $z$ and the position of $Q$ is $Z$, then $Z=-mz/M=-qz/Q$ in order for the c.m. to be stationary. This would be the case for small oscillations of the particle about an equilibrium position, {\it i.e.} the normal modes of the system. The induced charge difference, with $q$ and $Q$ moving such that the c.m. is stationary, is
\begin{equation} 
\Delta q= \gamma_3 z^3 \left( 1-\frac{q^2}{Q^2} \right) \approx \gamma_3 z^3
\end{equation}  
Again the signal is induced by the cubic term, which gives rise to signals at $\omega_z$ and $3 \omega_z$. If the oscillation frequency of $q$ in the environment $Q$ is reduced from the single-particle value, then this shifted frequency will be seen in the signal around $\omega_z$, so long as the term in $\gamma_3$ is present.

If the amplitudes of motion are larger than the interparticle spacing, the normal mode description no longer applies. In this case the motion of an individual particle does not necessarily give rise to a compensating motion of the rest of the cloud. Therefore, the individual oscillation frequencies of all particles are observed directly in the detected signal \cite{WER05}, {\it i.e.} via the non-linear term in equation (32).

\section{Resistive cooling of the axial motion}
For a single ion, the current $I$ running through the resonant RLC-circuit results in an exponential decrease of the particle's kinetic energy, which is given by \cite{WIN75}
\begin{equation} 
E(t)=E_0 \mbox{exp} (-t/\tau) \quad \textrm{and} \quad \tau=\frac{m (2z_0)^2}{R q^2}
\end{equation}
where $R$ is the resistance of the RLC-circuit. In resonance, the impedance of an RLC-circuit is real and acts as an ohmic resistor with resistance $R=QL\omega_z$, where $Q$ is the quality factor of the circuit and $L$ the inductance. The bandwidth is given by $\Delta \nu_z=\nu_z/Q$ and for typical values ($\nu_z \sim 1$MHz, $Q \sim 1000$) the 
bandwidth is $\sim 1$ kHz. The resistive cooling time constant, expressed in terms of the circuit components, is given by
\begin{equation} 
\tau=\frac{4(z_0 \sqrt{m})^3}{LQ\sqrt{q^5U_0}}
\end{equation}
The cooling process finally leads to an equilibrium between the particle's kinetic energy and the temperature of the heat bath. However, the noise temperature in the electronics may exceed the physical temperature of the RLC-circuit, leading to somewhat higher temperatures of the particles \cite{DJE04}. Deviations from the exponential cooling behaviour may occur when the dissipated power is not proportional to the kinetic energy of the particles. For example, if the trap is anharmonic, $\omega_z$ depends on the particle's kinetic energy and may move out of resonance with the RLC-circuit \cite{BRO86,MAJ04}.

For coherent c.m.-motion of a particle cloud with $N$ particles, the total induced current is given by $I_{tot}=NI$, and correspondingly this motion is strongly cooled. Apart from the coherent c.m.-motion, there can be non-coherent motions of ions within the cloud. As described above, these individual ion motions give rise to a resultant induced signal in the RLC-circuit \cite{MAJ04,WER05}. Since the signals from the non-coherent motions cancel to first order, the corresponding cooling is weaker and therefore the cooling times are longer \cite{HAF03}. 

As discussed in section 4.2, space charge effects generally lead to a shifted and broadened distribution of axial frequencies of trapped particles. Resistive cooling is only effective within the bandwidth of the external circuit, which may be less than the range of frequencies expected from the space charge effects. In large ion clouds the axial frequency distribution may be considerably broadened by space charge effects (see {\it e.g.} inset figure~\ref{fig4}). In that case, it is expected that the inter-particle collisions continuously thermalise the cloud by redistributing energy over the cloud's axial frequencies so that the resistive cooling process continues, but at a reduced rate limited by the thermalisation process. The inter-particle collision rate for a single component, trapped particle cloud is characterised by $k \propto q^4 m^{-1/2} n T^{-3/2}$ where $q$ is the particle's charge, $m$ its mass, $n$ the particle number density and $T$ the intrinsic temperature of the ion cloud \cite{DIV01}. Therefore, thermalisation is most efficient for light, highly charged particles in a dense, cold ion cloud. It may be expected that, for an efficient thermalisation process, the cooling time of a particle cloud scales with the bandwidth of the RLC-circuit. Therefore, in practical applications, a compromise has to be found between a sufficiently small cooling time constant $\tau \propto 1/Q$ and a sufficiently large bandwidth $\delta \nu_z \propto 1/Q$.

\section{Conclusions}
We have presented a thorough analysis of the formation of image charges. Such a study is relevant because the images formed in the trap electrodes of a Penning trap will be picked up by an electronic detection circuit attached to the trap electrodes. Our analysis was done for the case of two parallel conducting plates and for the case of two hollow conducting spheres. It was found that in both systems mainly the negative (odd) images contribute to the induced charge difference $\Delta q$ between the two electrodes. These images lead to (higher order) odd dependences of $\Delta q$ on the particle's position $z$ between the electrodes and its oscillation frequency $\omega_z$. We therefore concluded that an electronic detection system can, in principle, pick up higher order frequencies of the particle motion as well. Such frequencies can, for example, be induced by individual particle motions at a frequency $3 \omega_z$, rather than exclusively at the common centre-of-mass motion at the frequency $\omega_z$.

We have also derived analytic expressions for the shifted frequencies of trapped particles, which are due to image charge effects and space charge effects. Both effects lead to a downwards shift of the axial frequency $\omega_z$ and the modified cyclotron frequency $\omega_+$, but to an upwards shift of the magnetron frequency $\omega_-$. These shifts depend strongly on the charge density of the confined particle cloud and on geometrical trap parameters. Our simulations qualitatively confirmed the downwards shift of the axial frequency. In addition, they also showed that in cold ion clouds, with motional amplitudes smaller than the cloud dimensions, the frequency shift depends on the particle's average distance from the cloud centre.

Resistive cooling of a trapped particle cloud depends also on the image charges induced in trap electrodes, and on the energy dissipation in an external heat bath formed by a frequency resonant RLC-circuit. Due to the limited bandwidth of such a circuit, space charge induced shifts and broadening of the axial trapping frequency distribution need to be taken into account. Furthermore, to first order only the centre-of-mass mode induces a signal in the circuit. Since the cooling is effective only within the bandwidth of the circuit, resistive cooling of large clouds, where considerable space charge effects occur, may differ significantly from the single particle case and may lead to much longer cooling times.

\ack{
We acknowledge discussions with G.H.C. New. This work was supported by the European Commission within the framework of the HITRAP project (grant no. HPRI-CT-2001-50036).}


\section*{References}
\begin{thebibliography}{99}
\bibitem{BRA99} Bradley M, Porto J V, Rainville S, Thompson J K and Pritchard D E 1999 {\it Phys. Rev. Lett.} {\bf 83} 4510
\bibitem{FAR95} Farnham D L, Van Dyck Jr. R S and Schwinberg P B 1995 {\it Phys. Rev. Lett.} {\bf 75} 3598
\bibitem{VAN89} Van Dyck Jr R S, Moore F L, Farnham D L and Schwinberg P B 1989 {\it Phys. Rev. A} {\bf 40} 6308
\bibitem{DHO95} Dholakia K, Horvath G Zs K, Power W, Segal D M and Thompson R C 1995 {\it Appl. Phys. B} {\bf 60} 375
\bibitem{THO04} Thompson J K, Rainville S and Prichard D E 2004 {\it Nature} {\bf 430} 58 
\bibitem{VAN98} Van Dyck Jr R S, Farnham D L, Zafonte S L and Schwinberg P B 1998 {\it Trapped charged particles 
and fundamental physics} ({\it AIP conference proceedings} vol 457) ed D H E Dubin and D Schneider (New York: AIP)
\bibitem{BRO86} Brown L S and Gabrielse G 1986 {\it Rev. Mod. Phys.} {\bf 58} 233
\bibitem{GHO95} Ghosh P K 1995 \textit{Ion Traps} (Oxford: Clarendon Press)
\bibitem{KRE91} Kretzschmar M 1991 {\it Eur. J. Phys.} {\bf 12} 240
\bibitem{TIN01} Tinkle M D and Barlow S E 2001 {\it J. Appl. Phys.} {\bf 90} 1612
\bibitem{FEY65} Feynman R P, Leighton R B and Sands M 1965 {\it The Feynman Lectures on Physics Vol. II} (Reading: Addison-Wesley)
\bibitem{JAC75} Jackson J D 1975 {\it Classical Electromagnetism} (New York: Wiley)
\bibitem{BLE65} Bleaney B I and Bleaney B 1965 \textit{Electricity and Magnetism 2$^{nd}$ edition} (Oxford: Oxford University Press)
\bibitem{WIN75} Wineland D J and Dehmelt H G 1975 {\it J. Appl. Phys.} {\bf 46} 919
\bibitem{POR01} Porto J V 2001 {\it Phys. Rev. A} {\bf 64} 023403
\bibitem{JEF83} Jeffries J B, Barlow S E and Dunn G H 1983 {\it Int. J. Mass Spectrom. Ion Proc.} {\bf 54} 169
\bibitem{MAJ04} Major F G, Gheorghe V N and Werth G 2005 \textit{Charged Particle Traps} (Berlin: Springer-Verlag)
\bibitem{MEI88} Meis C, Desaintfuscien M, and Jardino M 1988 {\it Appl. Phys. B} {\bf 45} 59
\bibitem{GAB89} Gabrielse G, Haarsma L and Rolston S L 1989 {\it Int. J. Mass Spectr. Ion Proc.} {\bf 88} 319
\bibitem{WIN05} Winters D F A, Abdulla A M, Castrej\'on-Pita J R, de Lange A, Segal D M and Thompson R C 2005 {\it Nucl. Instr. Meth. Phys. Res. B} {\bf 235} 201
\bibitem{VOG05} Vogel M, Winters D F A, Segal D M and Thompson R C 2005 {\it Rev. Sci. Instrum.} {\bf 76} 103102
\bibitem{WER05} Werth G 2005 {\it Private communication}
\bibitem{DJE04} Djekic S, Alonso J, Kluge H J, Quint W, Stahl S, Valenzuela T, Verd\'u J, Vogel M and Werth G 2004 {\it Eur. Phys. J. D} {\bf 31} 451
\bibitem{HAF03} H\"affner H, Beier T, Djekic S, Hermanspahn N, Kluge H J, Quint W, Stahl S, Verd\'u J,
Valenzuela T and Werth G 2003 {\it Eur. Phys. J. D} {\bf 22} 163
\bibitem{DIV01} Diver D 2001 {\it A Plasma Formulary for Physics, Technology and Astrophysics} (Berlin: Wiley)
\end{thebibliography}
\end{document}